\documentclass[12pt]{article}
\usepackage{graphicx}
\usepackage{amsmath} 
\usepackage{amssymb} 
\usepackage{amsfonts} 
\usepackage{rotating}
\usepackage{subfigure}

\begin{document}



\title{A biclustering approach to university performances: an Italian case study}

\author{Valentina Raponi\footnote{Dipartimento di Scienze Statistiche, Sapienza Universit\`{a} di Roma, Roma, Italy}, Francesca Martella\footnote{Dipartimento di Scienze Statistiche, Sapienza Universit\`{a} di Roma, Roma, Italy} and Antonello Maruotti\footnote{Southampton Statistical Sciences Research Institute, Southampton, UK}}

\maketitle

\begin{abstract}
University evaluation is a topic of increasing concern in Italy as well as in other countries. In empirical analysis, university activities and performances are generally measured by means of indicator variables, summarizing the available information under different perspectives. In this paper, we argue that the evaluation process is a complex issue that can not be addressed by a simple descriptive approach and thus association between indicators and similarities among the observed universities should be accounted for. Particularly, we examine faculty-level data collected from different sources, covering 55 Italian Economics faculties in the academic year 2009/2010. Making use of  a clustering framework, we introduce a biclustering model that accounts for both homogeneity/heterogeneity among faculties and correlations between indicators. Our results show that there are two substantial  different performances between universities which can be strictly related to the nature of the institutions, namely the \textit{Private} and \textit{Public} profiles . Each of the two groups has its own peculiar features and its own group-specific list of priorities, strengths and weaknesses. Thus, we suggest that caution should be used in interpreting standard university rankings as they generally do not account for the complex structure of the data.\\
{\bf Keywords:} Biclustering; University performance; Gaussian mixture; Factor model
\end{abstract}

\section{Introduction}
\label{s:intro}
Measuring performance in higher education has become an important issue in OECD countries (OECD, 2013; Stolz et al., 2010). There is a permanent search for lodestone of academic quality and prestige, which contributes to increase the controversy that fuels the university ranking industry. Most of the existing empirical works on universities evaluation look at student-level administrative data (Belloc et al., 2012; Belloc et al., 2011; Belloc et al., 2010; Bini et al., 2009; Arulampalam et al., 2004; Biggeri et al., 2001) aiming at capturing students performance  or at aggregate data aiming at comparing universities under different perspectives (Triventi and Trivellato, 2009; Rampichini et al., 2004; Abbott and Doucouliagos, 2003). While defining the output of student-level analyses is straightforward (e.g. dropout), a major hurdle in evaluating higher education institutions relies on the definition of appropriate indicators able to measure the {\it quality} and the {\it effectiveness} of the university activities. Nowadays a unifying
proposal of university performance indexes is still more desirable since tools for information about university performance are
more easily available than in the past. As far as we know there are no standard rules to compute such indexes and their properties have been widely debated over many years. Although the use of descriptive and synthetic indicators is crucial to inform the non-specialist people (see e.g. Censis 2012), purely descriptive approaches may fail in capturing the complex university structure. It could be stressed that two main drawbacks should be avoided: ambiguity, which occurs when the global index signals a bad
situation but the sub-indexes do not, and eclipsicity, which indicates good general conditions while the
sub-indexes say the contrary. Thus, evaluating university performance on a single indicator may be inappropriate because it may ignore the multi-factor dimension of performance.

A multivariate analysis, which accounts for the several university aspects (namely {\it productivity}, {\it teaching}, {\it fund raising and research} and {\it internationalization}), is crucial to avoid misleading results, and more complex statistical approaches can be introduced to {\it measure} university performances. Interesting proposals have been introduced in a clustering framework (Ibanez et al., 2014; Valadkhani and Ville, 2009). With the aim of identifying similarities and/or differences among universities, by claiming the existence of groups with close characteristics, we aim at extending this branch of literature by jointly clustering universities and performance indicators, leading to the identification of different {\it double} partitions. A possible and straightforward approach is to separately cluster universities and indicators. Nevertheless, it is widely known that this approach does not allow nor to specify an overall objective function (thus lucking in optimality properties) nor to take into account the dependence structure between rows and columns of the considered data matrix. Following an idea that dates back to Fisher (1969) and Hartigan (1972), one may find more appropriate to perform a simultaneous clustering of both units (i.e. universities) and variables (i.e. indicators). This is generally called biclustering  but also known under a broad range of different names, including double clustering, block clustering, bidimensional clustering, co-clustering, simultaneous clustering and block modeling (for a review see e.g. Van Mechelen et al., 2004; Madeira and Oliveira, 2004). It is worth to note that the use of this methodology is often needed since standard cluster analysis followed by a factorial reduction (see e.g. Fraley and Raftery, 2002) may fail in detecting relevant information in the data. In particular, in the context of university performance analysis, these standard methods may lead to a significant misinterpretation of the results since many performance patterns are common to groups of universities only under a specific set of indicators. Therefore, biclustering allows to achieve the further goal of detecting groups of universities with similar behavior
characterizing a specific subset of indicators.

 Empirical analyses based on aggregate data often ignore the varied performances which occur within universities at disciplinary level, and this may bias the results. Comparatively, little analysis of performance has been conducted at the disciplinary level, largely focused on compiling rankings of journals and of departments according to their productivity (Smyth and
Smyth, 2001; Pomfret and Wang, 2003; Macri and Sinha, 2006). However, we believe that the best way to compare faculties, universities, etc., is to focus on a particular research field and to use clustering techniques rather than straight ranking. To avoid case-mix problems, we focus only on Economics faculties and collect information on several aspects of their activities in such a way to create a large set of indicators able to capture different dimensions of faculty activity. We introduce a set of 24 indicators containing information about teaching, productivity, research and internationalization indicators. Such indicators are measured on 55 Italian Economics Faculties referring to the academic years 2008-2009 and 2009-2010. To get a picture of the overall aspects involved in the higher education system, data are collected from different sources. Mainly, we used data from the Ministry of University
and Research (MIUR), the National Committee for the Evaluation of the University System (CNVSU), CINECA and the Lifelong Learning Programme. Data on the Erasmus project are provided by the Erasmus and International Relationship Offices of each University.
In some cases the indicators are calculated averaging over a period based on the last available academic years (2008-2009, 2009-2010). The main reason of this choice is making the indicators as stable (and unbiased) as possible over time and thus not affected by possible errors in data transmission or other occasional events.
Italy represents an interesting case study since its university system in last years has developed a high-level of decentralization and has reached an increased autonomy in managing and allocating resources. Therefore a need of
performance measurement system has been claimed both by central government and university.

The plan of the paper is as follow. In Section 2, we describe the data, introduce the indicators and provide summary (descriptive) statistics. The model-based biclustering used for the analysis has been proposed by Martella et al. (2008) and described in Section 3, along with computational details needed to obtain parameter estimates. Results are discussed in Section 4, whereas Section 5 provides conclusions and future development.

\section{Data description}

\subsection{Description of the variables}
We look at four major aspects of university performance: productivity, teaching, research and internationalization. Each of these four categories will be described in depth in the following subsections
\subsubsection{Productivity indicators}
There is not a unique measure of the concept of productivity. Faculties serve multiple objectives and their operation can be assessed in term of effectiveness of achieving various objectives.
To measure the productivity of a faculty, defined as a ratio of output to input, we must somehow evaluate its output.
In this sense, we propose the following indicators.
\begin{description}\item[P1 - Rate of persistence between the first and the second academic year:]
By looking at a specific cohort, the indicator represents the number of students enrolled in the second year, among those matriculated in the previous academic year, over the number of students matriculated in the previous academic year. \\The index has a higher value for the faculties with a higher transition rate from the first to the second year of study.

\item[P2 - Achieved credits:]
Credits achieved by all students during the last academic year/ (enrolled students*60).\\The quantity in the denominator represents the maximum amount of credits achievable during an academic year. The index measures the amount of credits actually achieved by the enrolled students over the maximum amount of credits achievable during the considered academic years.

\item[P3a - Rate of regular students enrolled in the 3-year bachelor-level Programs:] The index measure the portion of regular students in the 3-year bachelor programs with respect to all the enrolled students. Formally it is defined as the number of students enrolled in a 3-year  bachelor-level Programs (or in older system Programs) for a number of years not exceeding the official length of considered Program and net for freshman students/ total number of students enrolled in the 3-year bachelor-level Programs (or in older system Programs) net for freshmen,  students who already have a degree and students with unknown year of first registration.\\ Following the definition adopted by the MIUR, a regular student is a student enrolled in the university system for a number of years not exceeding the official length of considered Program.

\item[P3.b - Rate of “regular students” enrolled in the 2-year master-level Programs:]
Students enrolled in a 2-year master-level Program for a number of years not exceeding the official length of considered Program and net for freshmen / total amount of students enrolled in the 2-year master-level Programs and net for freshmen, for  students who already have a degree and for students with unknown year of first registration.\\ The index measures the portion of “regular students” in the 2-year  master-level Programs with respect to all the enrolled students in the analyzed programs.

\item[P4.a – Rate of  “regular graduate-students”  in the 3-year bachelor-level Programs:]
Students graduated in time in a 3-year bachelor-level Program (or in an older system Program) / Total amount of students graduated in the 3-year bachelor-level Programs (or in older  system Programs) net for “early-graduated” students, for  students who already have a degree and for students with unknown year of first registration.\\ Following the definition adopted by the MIUR, a “early-graduated” student is a student graduated before the end of the official length of the considered Program. The variable represents the portion of students graduated in time in the 3-year bachelor-level Programs.

\item[P4.b - Rate of  “regular graduate-students”  in the 2-year master-level Programs:]
Students graduated in time in a 2-year master-level Program / Total number of students graduated the 2-year master-level Programs net for “early-graduated” students, for students who already have a degree and for students with unknown year of first registration. \\ The indicator represents the portion of students graduated in time in the 2-year master-level Programs.
\end{description}
\subsubsection{Teaching indicators}
Teaching is a crucial activity in the Italian university system and often it is the main activity of university staff.
Every aspect of teaching addresses the intellectual and personal development of our students and it represents an ongoing interaction with students through course design, teaching activities, assessment and feedback. Thus, measuring available {\it human capital} and resources is extremely important in order to offer an appropriate service to students. Therefore, we suggest the indicators listed below, rewarding the universities with the highest values of these variables.
\begin{description}
\item[D1 - Permanent professors per credits:]
Permanent professors in the last two calendar years / total amount of credits taught  by permanent professors during their teaching activities in the last two calendar years.\\  Being fixed the amount of credits, the indicator achieves a higher value when the number of permanent professors is higher. In other words, the indicator rewards faculties in which the amount of credits is provided by a higher number of permanent professors.

\item[D2 - Permanent professors per enrolled student:]
Permanent professors in the last two calendar years /  total number of enrolled students in the last two academic years.

\item[D3 - Seats per enrolled student in the academic year 2009-2010:]
Number of total seats in the last academic year / total amount of students enrolled in the last academic year.

\item[D4 - Seats per student enrolled in the academic year 2008-2009:]
Number of total seats in the academic year 2008-2009 / total amount of students enrolled in the academic year 2008-2009.

\item[D5 - Researchers to professors ratio:]
Researchers available in the last two academic years / ordinary professors in the last two academic years.

\item[D6 -  Monitored teaching activities:]
Number of monitored teaching activities / total amount of available teaching activities
\end{description}

\subsubsection{Research indicators}

Nowadays, central resources are poor and universities are looking for research funds more often than before. Performing high-level research attracts funds and may contribute to the development of the structure. Therefore, we select the following indicators
\begin{description}
\item[R1 - Financed research units per professor:]
Total amount of national and local research units financed by the PRIN Program in the last three years/  average number of permanent professors during the last three calendar years.
\item[R2 - Average funding per research unit:]
Total amount of funding obtained by national and local research units from the participation in the PRIN program /  total number of financed units.
\item[R3 – Submitted Projects per professor:]
Total number of research units submitted for co-financing concerning the PRIN Program during the last three calendar years/ average number of permanent professors during the last three calendar years.
\item[R4 - Success rate in the PRIN Program:]
Total amount of research units (national and local) financed by the PRIN program/ total number of units submitted for co-financing.
\item[R5 - Average funding for international research per professor:]
Total amount of funding from the European Union and other foreign public/private institutions and projects with high scientific relevance financed by MAE or MIUR  in the last three calendar years /  average number of permanent professors in the last three calendar years.
\item[R6 - Financed research projects per professor:]
Total number of research units financed by the European Commission in the last three calendar years / average number of permanent professor in the last three  calendar years.
\item[R7 - Average funding for FIRB project:]
Funding for FIRB Project obtained during the last three calendar years / total number of financed projects.
\end{description}

\subsubsection{Internationalization indicators}
Internationalization of education and students mobility represent top priorities to be a prestigious university. The aim is to promote cooperation between higher education institutions and contribute to the development of a pool of well-qualified, open-minded and internationally experienced young people as future professionals. We want to measure this capability by examining the following indicators.
\begin{description}
\item[I1 – Outgoing student mobility:]
Number of students who completed a period of study abroad through the ERASMUS Project or other similar projects in the last two academic years/ total number of enrolled students (net of freshmen)  in the last two academic years.
\item[I2 – Incoming student mobility:]
Foreign students with ERASMUS scholarship during the last two academic years / Total number of enrolled students in the last two academic years.
\item[I3 - Host Universities:]
Number of foreign universities that hosted ERASMUS students in the last two academic years  / Total number of permanent professors in the last two calendar years.
\item[I4 - International Opportunities:]
Number of funding obtained due to international cooperation activities in the last three calendar years/ average number of permanent professors during the last three calendar years.
\item[I5 - Courses with double or joint title:]
 Number of Programs with double or joint title  / total number of activated Programs.
\end{description}

\subsection{Sample characteristics}
\begin{figure}
\centering
\subfigure[Productivity]{\includegraphics[scale=.4]{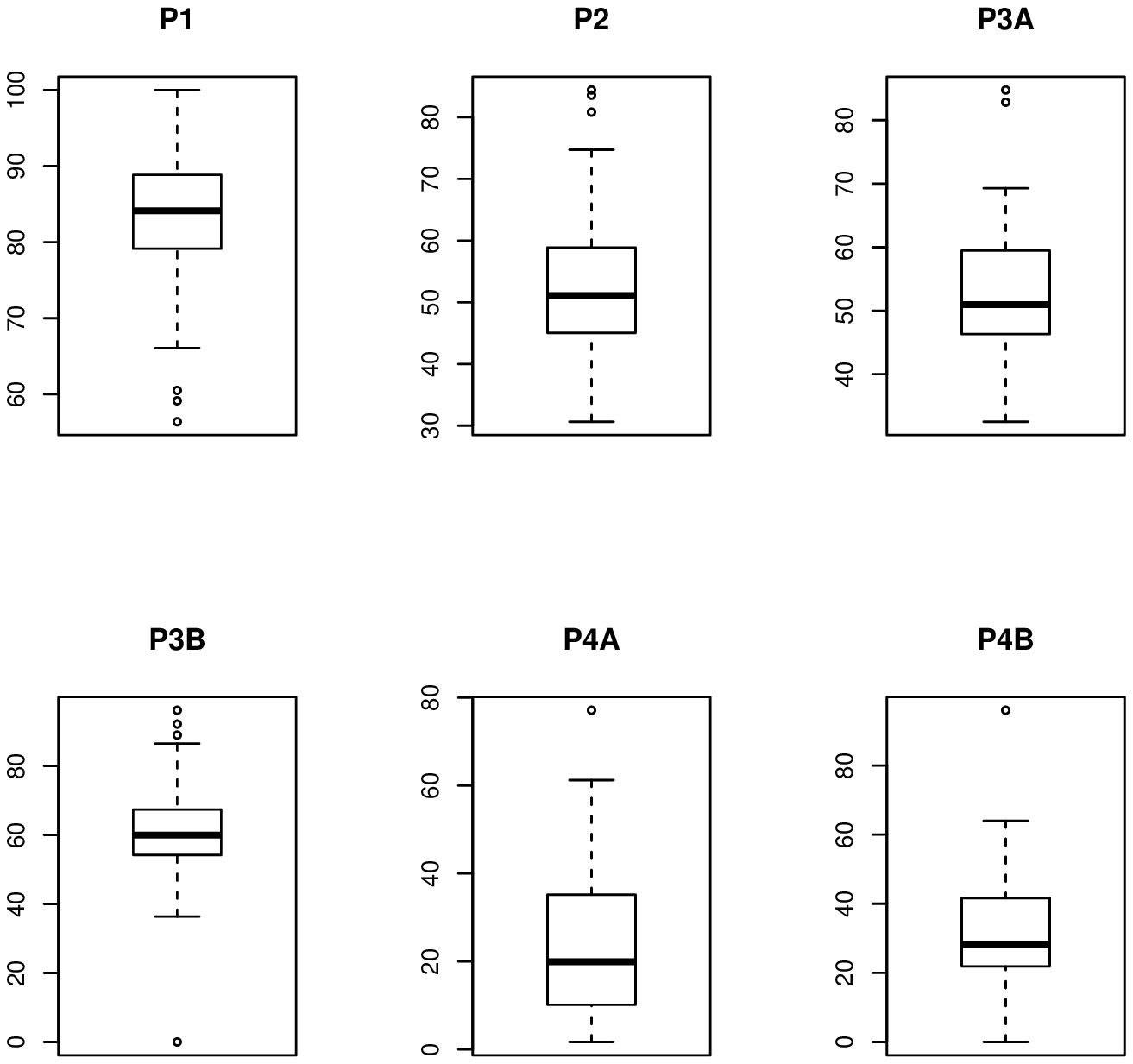}}\qquad
\subfigure[Teaching]{\includegraphics[scale=.4]{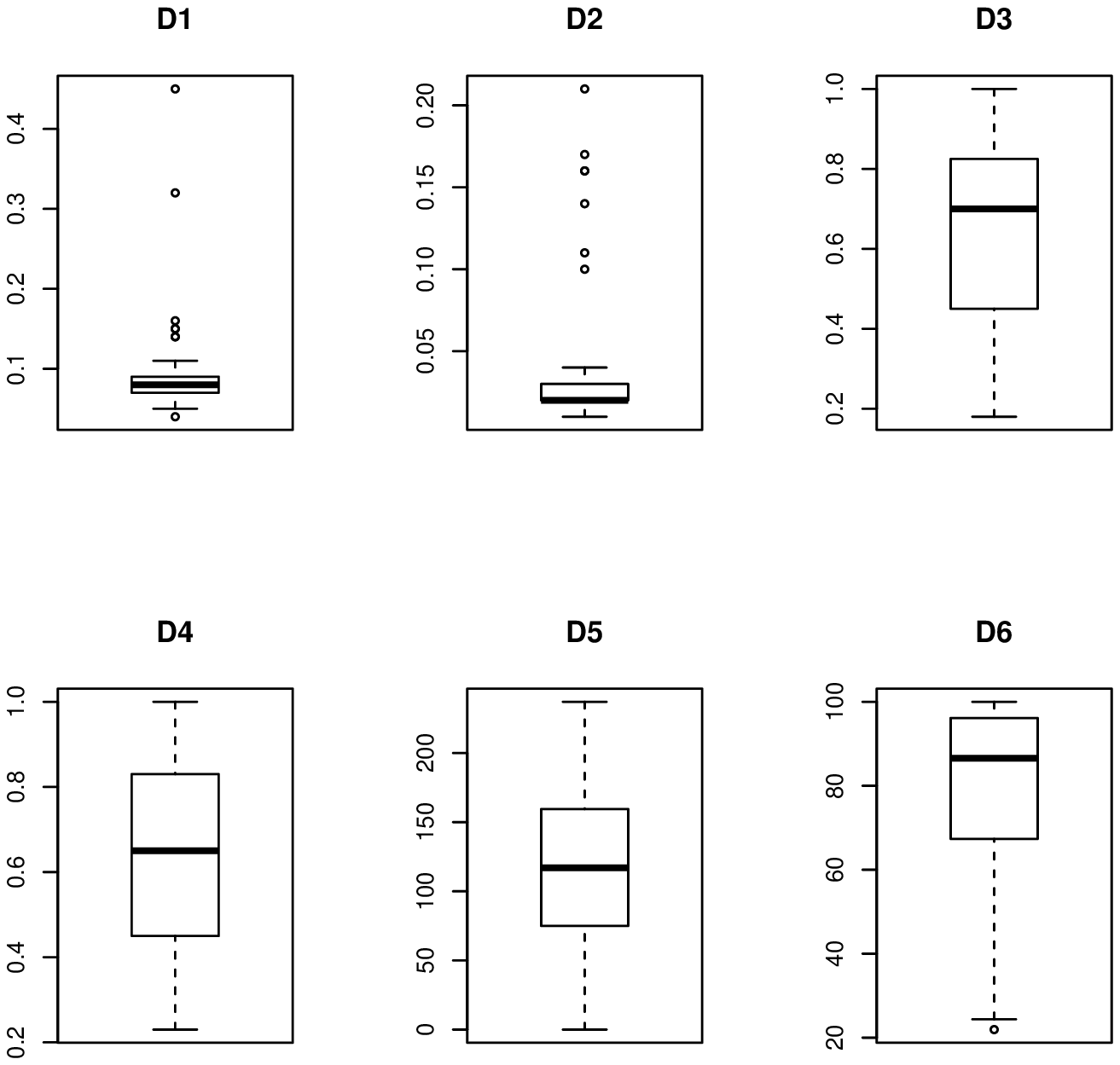}}\\
\subfigure[Research]{\includegraphics[scale=.4]{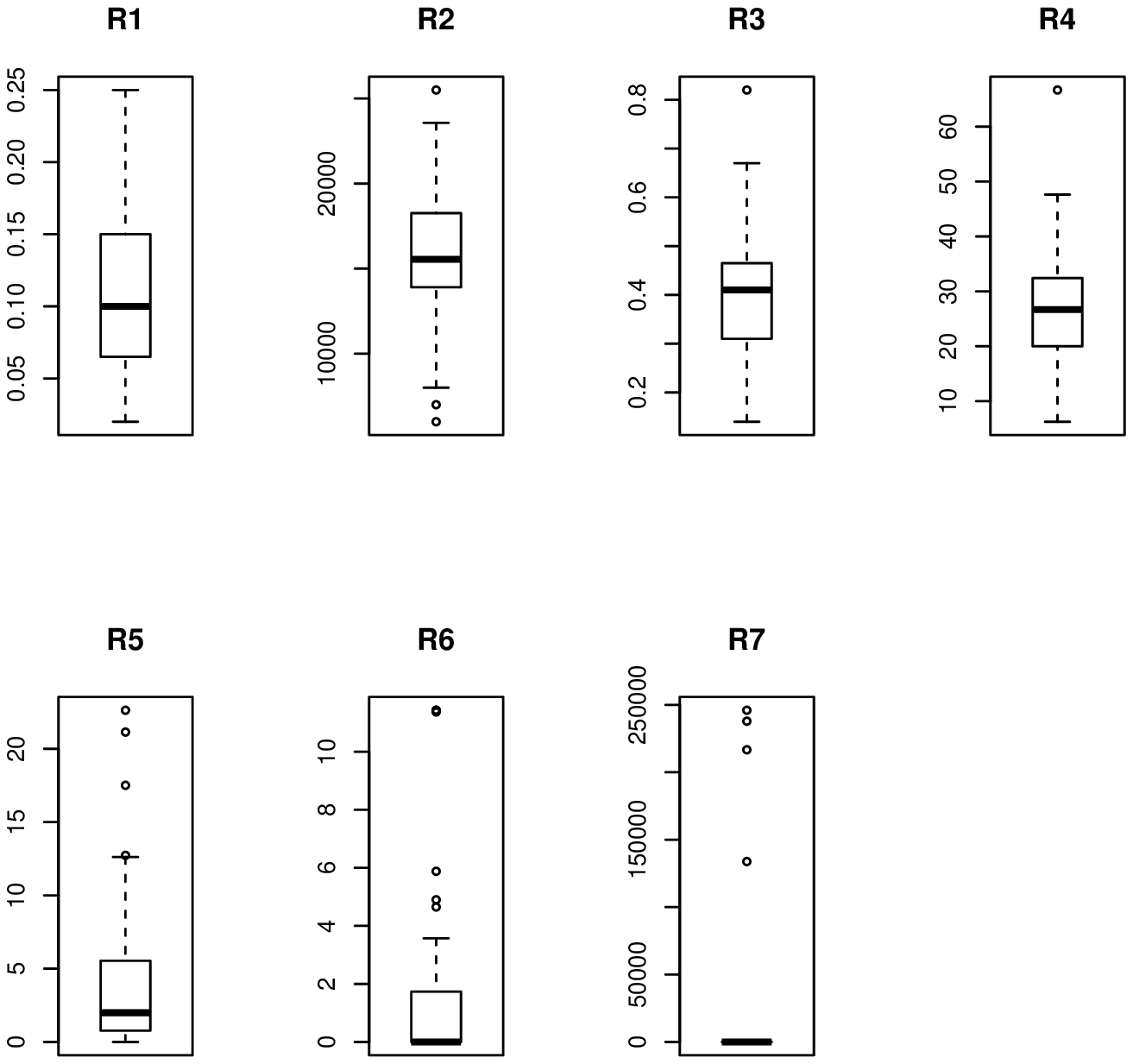}}\qquad
\subfigure[Internationalization]{\includegraphics[scale=.4]{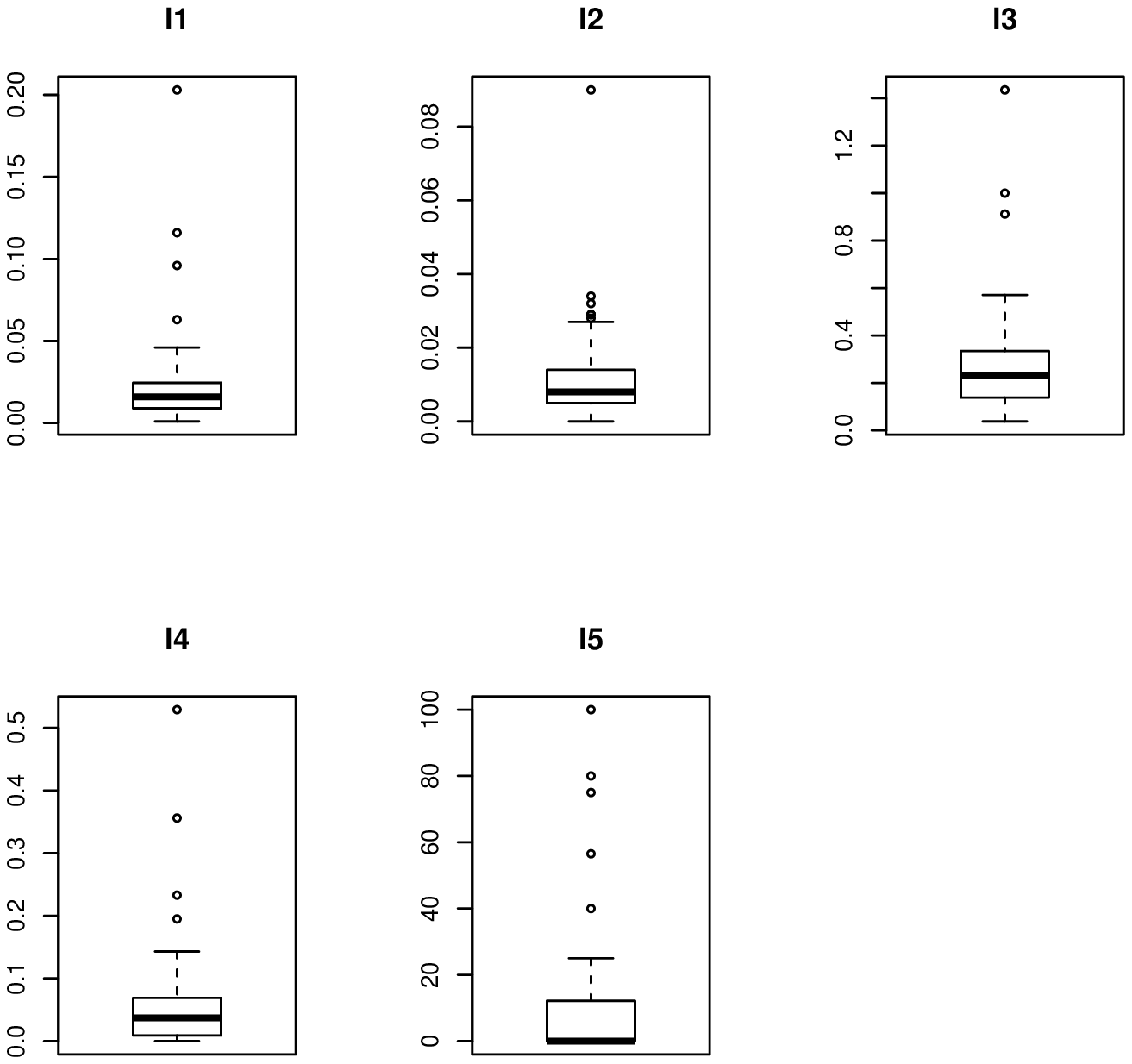}}
\caption{Summary statistics}
\label{summary}
\end{figure}

Our final data set consists of 55 Economics Faculties, over the 60 currently belonging to the Italian University system. The 5 ones not included in the sample have been dropped due to the lack of information on one or more of the introduced indicators. In particular, we collected data on 48 public and 7 private institutions. Summary information on all the considered indicators are provided in Figure \ref{summary}.
Most of the variables show high variability, providing evidence of heterogeneity in the sample. Most of the considered units lacks in internationalization with few exceptions only, which can be considered as outliers (in a broad sense). Similarly, on the research field, it is possible to identify few units funded by national grants, whilst most of the considered Faculties do not success in any research grant.

\section{Model-based biclustering}
Our modelling framework is based on the biclustering approach proposed by Martella et al. (2008), where the idea is to approximate the data density by a mixture of Gaussian distributions with an appropriate component-specific covariance structure. More precisely, we consider a partition of indicators by imposing a binary and row stochastic matrix representing column partition, whereas the traditional mixture approach is used to defined university clustering.

\noindent Formally, let us define a $J$-dimensional vector, ${\bf y}_i$, representing the faculty profile for the $i$-th faculty over $J$ indicators ($i=1,2,\dots,I$). Conditional on the $k$-th component of the mixture model ($k=1,2,\dots,K$), ${\bf y}_i$ is specified as
\begin{equation}\label{eq1}
{\bf y}_{ik} = \boldsymbol{\mu}_k+{\bf B}_k{\bf u}_{k}+{\bf e}_{ik}
\end{equation}

\noindent where $\boldsymbol{\mu}_k$ is the $J$-dimensional component-specific mean vector, ${\bf B}_k=\{b_{jlk}\}$ $(j=1,\dots,J; l=1,\dots,L_k; k = 1,\dots,K)$ is a binary row stochastic matrix representing column cluster membership, i.e. $b_{jlk} = 1$ if and only if the $j$-th indicator belongs to the $l$-th column cluster (and 0 otherwise), ${\bf u}_{k}$'s are  iid $L_k$-dimensional specific-block mean latent factors assumed to be drawn from $N({\bf 0},{\bf I}_{L_k})$, and ${\bf I}_{L_k}$ denotes the $L_k\times L_k$ identity matrix. Furthermore, ${\bf e}_{ik}$ are iid Gaussian component-specific error terms with mean ${\bf 0}$ and covariance matrix ${\bf D}_k={\rm diag}(\sigma_{1k}^2,\dots,\sigma_{Jk}^2)$, which are assumed to be independent of ${\bf u}_{ik}$.

\noindent Accordingly, the marginal density of ${\bf y}_i$ is as follows
\begin{equation}
f({\bf y}_i\mid \boldsymbol{\theta})= \sum_{k=1}^K\pi_kN_J({\bf y}_i;\boldsymbol{\mu}_k,\boldsymbol{\Sigma}_k)
\end{equation}

\noindent where $N_j(\cdot)$ represents the $J$-variate Gaussian distribution with component-specific $J$-dimensional mean vectors $\boldsymbol{\mu}_k$ and $J\times J$ component-specific covariance matrix $\boldsymbol{\Sigma}_k = {\bf B}_k{\bf B}_k'+{\bf D}_k$, $\pi_k$ are the prior probabilities of the mixture model, with $0\leq \pi_k\leq 1$ and $\sum_{k=1}^K\pi_k=1$, and $\boldsymbol{\theta}$ is a shorthand notation for all non redundant model parameters.

\noindent Note the proposed model may assume different specifications whether the
${\bf D}_k$ and ${\bf B}_k$ matrices are constrained to be equal across universities clusters or not.
The full range of possible constraints provides a class of 4 different models (see Table \ref{pgmm}).

\begin{table}
\begin{center}
\begin{tabular}{lcccc}
  Model & Membership & Error & Covariance & Total \\
& Matrix & Matrix & Parameters & Parameters\\
\hline
CC & Constrained & Constrained & $2J$ & $(K-1)+J (K+2)$\\
CU & Constrained & Unconstrained & $J (K+1)$ & $(K-1)+J(2K+1)$\\
UC & Unconstrained & Constrained & $J (K+1)$ & $(K-1)+J(2K+1)$\\
UU & Unconstrained & Unconstrained & $2KJ$ & $(K-1)+3KJ$\\

\end{tabular}\end{center}\caption{Covariance structures derived from different constraints}

\label{pgmm}
\end{table}

\noindent It is worth to note that the structure of  ${\bf B}_k$ leads to a peculiar form of the component-specific
covariance matrix of data. In fact, the adopted covariance model
 implies a block diagonal correlation structure, i.e. a block matrix having on its main diagonal $L_k$ blocks
formed by square matrices of size $J_l$ such that
the off-diagonal blocks are null matrices. In particular, the correlation
between variables depends on the variances only, in fact the smaller the
variance of one variable, the higher the correlation is among the other variables
in the block. Moreover, the fact that correlations depend on variances only is specific to biclustering, in which observations are assumed homogeneous (with small within variance) only under limited block of indicators that therefore are highly correlated. Finally, note that all variables have a
different variance, while the correlation between variables is equal to 1 if the
variables are within the same variable cluster, otherwise is equal to 0.
\noindent An alternating expectation conditional maximization (AECM) algorithm (Meng and van Dyk, 1997) is used for fitting these models. This algorithm is an extension of the EM algorithm that uses different specifications of missing data at each stage: when estimating $\pi_k$ and $\boldsymbol{\mu}_k$ the missing data are the unobserved component labels and, when estimating ${\bf B}_k$,  ${\bf u}_k$  and ${\bf D}_k$ the missing data are the component labels and the latent factors. In details, the AECM algorithm consists of the following steps:

\begin{enumerate}
\item Choose initial values for the parameter vector.
\item \textbf{First cycle}: at this stage,  we let ${\bf z} = \{{\bf z}_1,\dots,{\bf z}_n\}$ be the unobserved component labels, where $z_{ik} = 1$ if faculty $i$ belongs to component $k$ and $z_{ik}=0$ otherwise. Hence,
\begin{enumerate}
\item E-step: Update
\begin{equation}\label{post}
\hat{z}_{ik} = \frac{\pi_kN_J({\bf y}_i\mid\boldsymbol{\mu}_k,{\bf B}_k{\bf B}_k'+{\bf D}_k)}{\sum_{k=1}^K\pi_kN_J({\bf y}_i\mid\boldsymbol{\mu}_k,{\bf B}_k{\bf B}_k'+{\bf D}_k)}
\end{equation}
($i=1,...,n$ and $k=1,...,K$);
\item CM-step: Update $$\hat{\boldsymbol{\mu}}_k = \frac{\hat{z}_{ik}{\bf y}_i}{\sum_{i=1}^n\hat{z}_{ik}}\quad {\rm and}\quad \hat{\pi}_k=\frac{n_k}{n},$$ ($k=1,...,K$).
\end{enumerate}
\item \textbf{Second cycle}: at this stage,  we take the group labels ${\bf z}$ and the latent factors ${\bf u}$ to be the missing data. Hence,
\begin{enumerate}
\item E-step: Update $\hat{z}_{ik}$ as in (\ref{post}) ($i=1,...,n$ and $k=1,...,K$);
\item CM-step:
\begin{itemize}
\item Update  $\mathbf{B}_k$: we choose the unit value in each column as follows
\begin{equation*}
b_{jl}=\Biggr\{
\begin{array}{c l}
1 & \textrm{if} \quad \emph{H}_2(\cdot,b_{jl}=1)=
\max_{h}
 \emph{H}_2(\cdot,b_{jh}=1) \\
0 & \textrm{otherwise} \\
\end{array}
\end{equation*}
with $j=1,...,J$, $l,l,h=1,...,L_k$ and $l\neq h$; where $\emph{H}_2(\cdot,b_{jl}=1)$ is the expected complete-data log-likelihood given by
\begin{equation}
\begin{split}
\emph{H}_2(\mathbf{B}_k,\mathbf{D}_k,\mathbf{u}_k)=C+\sum_{k=1}^{K}\Biggr[\frac{n_k}{2}\log|\mathbf{D}^{-1}_k|-\frac{n_k}{2}tr\{\mathbf{D}^{-1}_k\mathbf{S}_k \}+\\+ \sum_{i=1}^{n}w_{ik}(\mathbf{y}_i-\mbox{\boldmath$\mu$}_k)\mathbf{D}^{-1}_k\mathbf{B}_kE(\mathbf{u}_k|\mathbf{y}_i,\mbox{\boldmath$\mu$}_k,\mathbf{D}_k,\mathbf{B}_k)-\\-\frac{1}{2}tr\Bigr\{\mathbf{B}_k'\mathbf{D}^{-1}_k\mathbf{B}_k\sum_{i=1}^{n}w_{ik}E(\mathbf{u}_k\mathbf{u}_k'|\mathbf{y}_i,\mbox{\boldmath$\mu$}_k,\mathbf{D}_k,\mathbf{B}_k)\bigl\}\Biggl]
,\end{split}
\end{equation}
with
\begin{equation*}
 \mathbf{S}_k
=\frac{\sum_{i=1}^{n}w_{ik}(\mathbf{y}_i-\mbox{\boldmath$\mu$}_k)(\mathbf{y}_i-\mbox{\boldmath$\mu$}_k)'}
{n_k}
\end{equation*}
and  $C$ is a normalizing constant independent of $\mathbf{u}_k$, $\mathbf{B}_k$ and $\mathbf{D}_k$.
\item Update $\mathbf{D}_k$:
\begin{equation*}
\hat{\mathbf{D}}_k=diag\{\mathbf{S}_k -\mathbf{B}_k\mathbf{L}_k\mathbf{S}_k\},
\end{equation*}
where $\mathbf{L}_k=\mathbf{B}_k'(\mathbf{B}_k\mathbf{B}_k'+\mathbf{D}_k)^{-1}$.
\item Update $\mathbf{u}_k$:
\begin{equation*}
\hat{\mathbf{u}}_{k}=\\E(\mathbf{u}_{k}|\mathbf{y}_{{i}}, z_{ik}=1)=\frac{
\mathbf{B}_k'(\mathbf{B}_k\mathbf{B}_k'+\mathbf{D}_k)^{-1}\sum_{i=1}^nw_{ik}(\mathbf{y}_{{i}}-\mbox{\boldmath$\mu$}_k)}{n_k}.
\end{equation*}.
\end{itemize}
\end{enumerate}
\item Compute the log-likelihood function for the current parameter values. If the function increase is larger than a fixed threshold, iterate once more according to 2. Otherwise, the process has converged.
\end{enumerate}

The AECM algorithm iteratively updates the parameters
until convergence to maximum likelihood estimates of the
parameters. The resulting $\hat{z}_{ik}$ values at convergence are estimates
of the a posteriori probability of group membership
for each observation and can be used to cluster universities
into groups while the $j$-th indicator
is allocated to $l$-th cluster through the matrix $\mathbf{B}_k$.

\subsection{Model selection criteria}
Biclustering model is a flexible and powerful approach to modeling data that are heterogeneous
and stems from multiple populations. It is well known that any continuous distribution can be approximated arbitrarily well by a mixture of normal densities (McLachlan and Peel, 2000). Nevertheless with too many components, the model may overfit the data and yield poor interpretations, while with too few components, the model may not be flexible enough to approximate the true underlying data structure. Hence, an important issue in clustering is the selection of the number of clusters. Most conventional methods for determining the number of clusters are based
on the likelihood function and some information criteria, such as Akaike information criterion (AIC) and Bayesian information criterion (BIC). These criteria would not underestimate the true number of clusters. Increasing the number of clusters always improves the fit of the model (as judged by the likelihood). But along with the improvement
comes an increase in the number of parameters, and the
improvement in fit has to be traded off against this increase. A criterion
for model selection is therefore needed.The AIC could be used
$$AIC = -2 \log \mathcal{L} + 2 \times \#par $$
where $\log \mathcal{L}$ is the log-likelihood of the fitted model and $\#par$ denotes the
number of parameters of the model. The first term is a measure of fit,
and decreases with increasing number of clusters. The second term is a
penalty term, and increases with increasing number of clusters. The BIC, which differs from AIC in the penalty term, can be also considered:
$$BIC = -2 \log \mathcal{L} + \#par \times \log n.$$

Compared to AIC, the penalty term of BIC has more weight in most applications; thus, the BIC often favours models
with fewer parameters than does the AIC. Although not considered herein, the use of Integrated completed likelihood as an alternative to the BIC often gave comparable clustering performance. Of course, several other criteria can be considered to perform model selection.

\section{Results}
\begin{figure}
\centering
\subfigure[Original data]{\includegraphics[scale=.5]{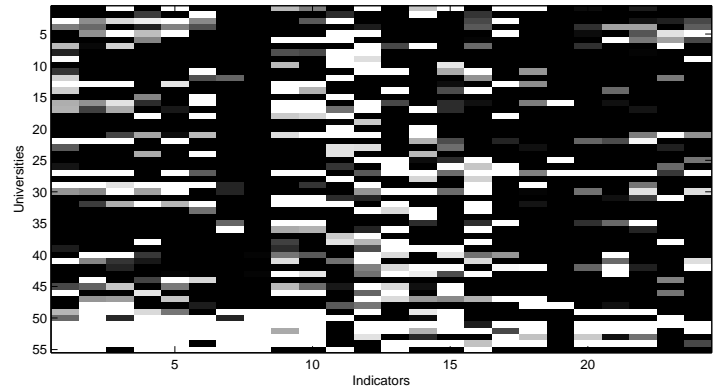}}\\
\subfigure[Clustered data]{\includegraphics[scale=.5]{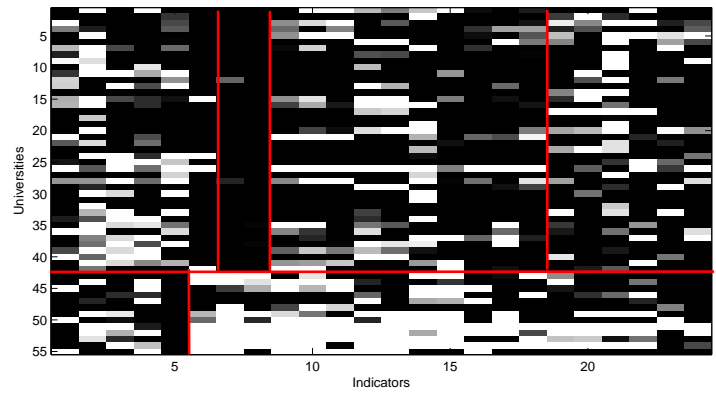}}
\caption{Original data and biclustering results}
\label{mod7}
\end{figure}

As mentioned in the introduction, our interest is to discover groups of Economic Faculties with similar behavior characterizing a specific subset of indicators in order to better capture the complex university performance structure. The proposed biclustering model was fitted to the described (standardize) data for different numbers of row and column clusters ($K=1,...,10$; $L_k=1,...,6$) and for different covariance structures (CC, CU, UC and UU). For each pair, we run the algorithm several times to avoid local maxima, choosing the best solution through the selection criteria described above. All of them agree in selecting the UU model with $K=2$ row-clusters and, for each of them, $L_1=4$ and $L_2=2$ column-specific clusters are respectively detected. The plot of the raw and ordered data is shown on Figure \ref{mod7}, where the red lines are used to separate row and column clusters.

In details, the two well-separated row (university) clusters have sizes 42 and 13, respectively and have been named {\it Public} and {\it Private} clusters. It is worth emphasizing that while the {\it Public} cluster consists only of public universities, few exceptions characterize the {\it Private} cluster:  6 public universities are more homogeneous with the private ones than with the rest of the public universities.
Table \ref{muk} shows the cluster-specific mean vectors  $\boldsymbol{\mu}=\{\boldsymbol{\mu}_k\}$  of standardized indicators. As can be easily seen, the {\it Private} group provides higher means for most of the indicators, as proof of better performances especially in terms of productivity and internationalization than {\it Public} universities. The situation is less clearcut if we look at teaching and research indicators. On average, {\it Private} profiles are very active on fund-raising (see R1, R3, R4, R6) and excel in obtaining funds from the European Union and other international institutions (see R5). Conversely, the {\it Public} group is characterized by a higher average funding resulting from the participation to national projects (PRIN and FIRB projects) as shown by the higher value in R2 and R7 indicators. With regard to the teaching activities, the {\it Private} profile puts more attention on the quality of their services available for each student, in term of both human and structure utilities (D1-D4), than the {\it Public} sector. The latter shows positive performance just in terms of the number of researchers and the percentage of monitored teaching activities.
On the other hand, Table \ref{uk} provides the partition of the considered indicators for both the {\it Public} and the {\it Private} group. We also provide the estimates for the latent factors (i.e. specific-block mean vectors) ${\bf u}=\{{\bf u}_k\}$, which represent deviations from the row cluster-specific mean vectors $\boldsymbol{\mu}=\{\boldsymbol{\mu}_k\}$. Then, for each of the two groups ({\it Public} and {\it Private}), the column specific clusters collect different characteristics of the universities belonging to the respective group, as it is also shown by the signs of latent factors. For the {\it Private} group, we obtain two column-specific clusters which can be easily interpreted as the \textit{strengths} and the \textit{weaknesses} of these institutions. The values of the specific-block mean estimates are quite far from each others (-0.49, 0.16) suggesting a good separation of the two  column clusters. In fact, the strengths of the private units are represented by the indicators listed under the column {\it Cluster 2} in Table \ref{uk} (which indeed show a positive latent factor mean equal to 0.16) and, consistently with the results in Table \ref{muk}, they coincide with the high performance in productivity, internationalization, more funding from European Union and international institutions and better services in terms of human and structure resources per student. An interesting partition is also obtained for the {\it Public} group. In this case we obtain four column clusters with a slightly worse but still good separation (specific-block mean estimates are 0.156, -0.004, 0.037, 0.099) characterized by interesting different university features. Cluster 1 shows the highest (and positive) value of the latent factors (0.156), meaning that features measured by indicators P1, D6, R1, R3, R4 and R7 contribute positively to increase the final output of the {\it Public} group. Such features are strictly connected to the intense research activity in national projects (mainly PRIN and FIRB) and to actions aimed at partner and take care of students during their course of study. Although with a smaller value than Cluster 1, both Cluster 3 and 4 as well make a positive contribution to the {\it Public} group's performance, with latent factors average equal to 0.037 and 0.099, respectively. In some sense, Cluster 3 tests the \textit{prestige} of the university. In fact, it mainly collects information about student's performance (in terms of number of achieved credits, regular students and students graduated on time in the bachelor programme), quality of institution's services and academic staff (indicators D4, D4, R2 and R6) and internationalization of education (measured by both the incoming and outgoing students' mobility and by the cooperation with other foreign higher education institutions).
Aspects related to M.Sc. degrees, as well as to job experiences and international opportunities are well defined in Cluster 4, which could be labeled as the level of \textit{international specialization} of students, university and academic staff.
Finally, although there is not a significant deviation from the row-cluster-specific mean in Table \ref{muk}, Cluster 2 focuses on the available \textit{human resources}, since the indicators D1 and D2 are clustered together.

\begin{table}
\begin{center}
\begin{tabular}{lcc|lcc}
 Variables & {\it Public} & {\it Private}& Variables & {\it Public} & {\it Private}\\
\hline
P1	&	-0.32	&	0.52	& D1	&	-0.32	&	0.97	\\
P2	&	-0.50	&	1.11	&D2	&	-0.38	&	1.12	\\
P3a	&	-0.38	&	0.67	&D3	&	-0.32	&	0.65	\\
P3b	&	-0.22	&	0.88	&D4	&	-0.35	&	0.70	\\
P4a	&	-0.50	&	0.91	&D5	&	0.25	&	-0.40	\\
P4b	&	-0.16	&	0.58	&D6	&	0.03	&	-0.37	\\
\hline
R1	&	-0.23	&	0.37	&I1	&	-0.32	&	0.64	\\
R2	&	-0.07	&	-0.22	&I2	&	-0.29	&	0.40	\\
R3	&	-0.18	&	0.25	&I3	&	-0.34	&	0.68	\\
R4	&	-0.17	&	0.32	&I4	&	-0.03	&	0.06	\\
R5	&	-0.22	&	0.51	&I5	&	-0.12	&	0.14	\\
R6	&	-0.21	&	0.52	\\
R7	&	0.04	&	-0.27	\\
\end{tabular}\end{center}\caption{Cluster-specific mean vectors}
\label{muk}
\end{table}

\begin{table}
\begin{center}
\begin{tabular}{cccc|cc}
\multicolumn{4}{c|}{\it Public} & \multicolumn{2}{c}{\it Private}\\
\hline
Cluster 1 & Cluster 2 & Cluster 3 & Cluster 4 & Cluster 1 & Cluster 2\\
\hline
P1 & D1 & P2 & P3b & D5 & P1\\
D6 & D2 &P3a& P3b  & D6 & P2\\
R1 &	& P4a & P4b & R2 & P3a\\
R3 & 	& D3 & D5 & R4 & P3b\\
R4 & 	& D4 & R5 & R7 & P4a\\
R7 &	& R2 & I4 & &P4b\\
 & & R6 & I6 & &R1\\
 & & I1 & & & R3\\
 & & I2 & & &R5\\
 & & I3 & & & R6\\
 & &  & & & I1\\
 & &  & & & I2\\
 & &  & & & I3\\
 & &  & & & I4\\
 & &  & & & I5\\
 & &  & & & D1\\
 & &  & & & D2\\
 & &  & & & D3\\
 & &  & & & D4\\
\hline
\multicolumn{6}{c}{Latent factors ${\bf u}$}\\
\hline
0.156 & -0.004 & 0.037 & 0.099 & -0.4953 & 0.165\\

\end{tabular}\end{center}\caption{Column-specific clustering}
\label{uk}
\end{table}

\section{Conclusions}
Performance measurement is defined as a process of quantifying the efficiency and the effectiveness of actions. It can be regarded as a first step for policy makers to ensure that university resources are properly allocated. This paper provides some insight into the development of multi-factor performance analysis by using a multivariate technique which accounts for several university features.

Under this perspective, we collected data on universities activity, creating an unique dataset. We simultaneously look at four major dimensions of universities activity, namely productivity, teaching, research and internationalization, with the further goal of measuring the (unobserved) {\it quality} of academic features, through the definition of latent factors. A common feature in most of the frameworks about the evaluation in higher education is the draft of a final ranking among universities with the aim of identifying top-universities and comparing their performances. Nevertheless, rankings may suffer from case-mix-related problems and final results may be significantly affected by the choice of the weights in averaging the indicators.

To avoid this kind of problems, we propose a biclustering-based approach with the aim of identifying homogeneous groups of universities sharing similar characteristics and attitude towards a specific set of indicators. Through an empirical application on 55 Italian Economics faculties, we jointly group both universities and performance indicators, leading to an intuitive and easily interpretable picture of the system. It highlights not only the strengths and weaknesses of each institution, but also clearly identify differences between public and private universities through the  different correlations with all the  aspects of institutions' activities. Thus, performance measurement on effectiveness and efficiency of academic activity is not simply a ranking list of different institutions; rather, it is a multi-dimensional framework able to capture the multi-output data structure. These results could help the policy makers to better understand {\it how} and {\it where} implement actions in order to both improve the weaknesses and strengthen the excellence.

It would be of interest for further research to look at the impact of adopted policies over time. This requires the extension of the so far introduced dataset, to include a time dimension, i.e. the creation of a panel dataset. Accordingly, biclustering should be extended to the analysis of three-way data, allowing to use such a powerful methodology to a wide range of real-world data.

\end{document}